# Simple graphene chemiresistors as pH sensors: fabrication and characterization

Nan Lei, Pengfei Li, Wei Xue, Jie Xu∗

*Mechanical Engineering, Washington State University, Vancouver, WA 98686, USA*

**Abstract:**

We report the fabrication and characterization of a simple gate-free graphene device as a pH sensor. The graphene sheets are made by mechanical exfoliation. Platinum contact electrodes are fabricated with a mask-free process using focused ion beam, and then expanded by silver paint. Annealing is used to improve the electrical contact. The experiment on the fabricated graphene device shows that the resistance of the device decreases linearly with increasing pH values (in the range of 4-10)in the surrounding liquid environment. The resolution achieved in our experiments is approximately 0.3 pH in alkali environment. The sensitivity of the device is calculated as approximately 2 k$\Omega$/pH. The simple configuration, miniaturized size and the integration ability make graphene-based sensors promising candidates for future micro/nano applications.

∗*Corresponding author: Jie Xu*
*Email: jie.xu@wsu.edu*
*Telephone: 360 546 9144*

# 1. Introduction

Graphene, as an ideal two-dimensional material, has rapidly received significant attention since its discovery in 2004 by Novoselov, Geim and co-workers [1]. It has been found that graphene has many unique electrical, mechanical and physical properties, such as massless Dirac quasiparticles[2], high carrier mobilities and capacities [1-6]. The Young's modulus of graphene is measured to be approximately 1 TPa, making it the strongest material ever measured [7].

The ability to precisely detect chemical and biological species is extremely important in many applications, such as genomics, clinical diagnosis and pharmaceutics. Traditional optical detection methods require very complex techniques and labeling processes. Alternatively, the electrical detection methods using novel nanomaterial devices have been extensively studied in the past decade [8-15]. As one of the most promising material, graphene has unique electronic structures: single atom thickness with the $\pi$ states (valance band) and $\pi^*$ states (conduction band) touching at the Dirac points. The symmetric band structure of graphene makes it directly amenable to chemical and physical modification. In addition, the high carrier mobility of graphene makes the modification detectable by simply monitoring its conductivity change. Since the discovery of graphene, there is great potential for building graphene-based high-sensitivity, label-free, miniaturized electrostatic or electrochemical sensors [16-18].

One of the key challenges in current research and development of graphene-based sensors is material handling and device fabrication [18]. Conventional microfabrication approaches, such as lithography and etching, often require multiple complex steps including masking and aligning. Also graphene is often configured as the semiconducting material in transistors, which require the fabrication of a gate. In this note, we report a simplified graphene device where the graphene is configured as a planar chemiresistor. The manufacture of this device consists of a mask-free focused ion beam (FIB) process and several easy post-processing steps that can be done manually. In the end, we demonstrate the ability of this device in monitoring pH values. As a new material, the application of graphene is still at the infant stage, and the performance of our graphene-based devices cannot yet compete against current commercial devices. However, the miniaturized size, low cost and the integration ability make graphene-based sensors more suitable for future micro/nano systems.

## 2. Materials and methods

Figure 1 sketches the entire mask-free fabrication process of our graphene sensor. Graphene sheets were first made by mechanical exfoliation of bulk graphite (Kish Graphite, purchased from Graphene Supermarket) with scotch tape and randomly deposited onto a 285 nm $SiO_2$ layer on a silicon wafer (P/B(100), 1-10 ohm-cm purchased fromuniversitywafer.com). This specific $SiO_2$ thickness is chosen because it can provide the best visibility of graphene sheets, which can be explained by Fresnel-law-based models as developed in [19, 20]. An optical microscope and a camera (Nikon MM-40 and DXM 1200) were then used to identify and locate individual graphene sheets (as sketched in Figure 1a). This is currently the easiest and most popular way to spot graphene crystallites among copious thicker flakes. Usually, a few micron-sized graphene can be found over a millimeter-sized area, which is already sufficient for our purpose, because only one graphene sheet is needed on each wafer piece for device fabrication.

A dual beam system (FEI Quanta 3D 200i) was used for the fabrication of electrodes and observation of the sample: an ion beam for electrode deposition and an electron beam for imaging. The area of interest on the sample was positioned at the eucentric point where the two beams cross, as shown in Figure 2a. At this location, the electron beam could provide *in situ* imagining of the graphene. The sample was tilted at 52 ° for optimized ion incident angle. The FIB employs highly-focused $Pt^+$ ion beam scanned over the desired areas of the sample surface. The Pt metalorganic precursor (methylcyclopentadienyltrimethyl platinum) is introduced by a gas-injection system during the ion beam scanning, providing the well-controlled ion-assisted deposition of the metal. Therefore, the Pt lines can be fabricated directly over the desired area of a graphene with approximately 20 nm precision.

In this study, to prevent the graphene from being damaged by the $Pt^+$ ions, two contact electrodes (10 μm×100 nm × 100 nm) were first deposited on two sizes of graphene under 16.0 kV acceleration voltage and 4.0 pA ion current. Then, another two rectangular electrodes (200μm × 10μm × 100 nm) were deposited to expand the contact electrodes under 30.0 kV acceleration voltage and 0.5 nA ion current. Figure 2b shows a microscopic image of the fabricated electrodes on both sides of a graphene sheet. To expand the contact platinum electrodes, two millimeter sized testing pads, as sketched in Figure 1c, were conveniently made by carefully painting a layer of silver on the wafer using high-purity conductive silver paint (SPI 05001-AB), operated under a stereo microscope (Nikon S2800). To eliminate the potential contamination of silver particles from dissipating into the testing samples, a layer of polydimethylsiloxane (PDMS) (Dow Corning Sylgard 184) was carefully cured on top of the silver electrodes and leaves a 200-300μm gap large enough for future experiments on the graphene sheet. The configuration of a fabricated graphene device is sketched in Figure 1d.

Previous studies show that the contact resistance of a metal-graphene interface can be affected by configuration [21] and temperature [22], and an annealing step can help improve the electrical interconnection [23]. Therefore, we processed our device with an annealing step at $200^{o}$C for 50 minutes. Figure 3 shows the I-V characteristics of a graphene device before and after annealing. We can see from Figure 3 that the I-V curve is symmetrical after annealing. This is due to the fact that our device has a symmetrical configuration consisting of two oppositely-connected Schottky diodes formed by graphene-metal conjunctions [24]. We suspect that the initial asymmetric I-V curve before annealing might be due to the bad electrode contact, which were improved during the annealing process.

## 3. Results and discussion

To perform pH sensing experiments, a wetting and drying process was used. After the resistance of the device in the air was stabilized, the first pH buffer drop was carefully placed on top of the graphene sheet as shown in Figure 4. After approximately 1~2 minutes when the resistance became stable, this pH buffer drop was carefully sucked up by vacuum, and then more buffer drops with different pH values were applied and removed repeatedly in the same way. In our experiments, buffers with pH values from 4 to 10 (standard buffer solutions from Fisher Scientific) were tested. During the experiment, the resistance of the device under a constant current were monitored and recorded in real time by a semiconductor analyzer (Agilent Technologies B1500A).

## 3.1 pH sensitivity

Figure 5a reports a real time measurement of the resistance of our graphene device under a constant current at 10 μA. The resistance of the device in the air is approximately 86 kΩ. When a drop of pH-4 buffer was placed on the graphene, the resistance rapidly decreased to 59.2 kΩ. The resistance is then decreased by 2.2 kΩ when pH-5 buffer was placed on the graphene. Similar amount of decrement is observed when the pH buffer was changed to pH-6, Ph-7, pH-8, pH-9 and pH-10. Figure 5b shows the average values (dots) and standard deviations (error bars) of the resistance from multiple measurements on this specific device. A linear correlation can be used to describe the curve in Figure 5b: $R$ = (-2.13 × pH + 66.11) kΩ. This equation indicates that the sensitivity of the particular device is 2.13 kΩ/pH.

To better characterize the grapheme-based sensor, we use the standard deviation to estimate the pH resolution using the following equation:

$$\Delta pH = \frac{\sigma_{max}}{k} \qquad (1)$$

Where $\Delta pH$ is the resolution of the grapheme device; $\sigma_{max}$ is the maximum standard deviation of resistance for multiple measurements from pH-4 to pH-10; $k$ is the sensitivity of a particular device. In Figure 5b, the standard deviations $\sigma$ at each pH value are shown below:

Table1: Standard deviations of the device under pH from 4 to 10.

| pH value | 4 | 5 | 6 | 7 | 8 | 9 | 10 |
|---|---|---|---|---|---|---|---|
| $\sigma$(k$\Omega$) | 2.06 | 1.63 | 1.75 | 0.89 | 0.13 | 0.71 | 0.36 |

The $\sigma_{max}$ in acid is 2.06 k$\Omega$, while the $\sigma_{max}$ in alkali is 0.71 k$\Omega$. For the particular device the sensitivity $k$ is 2.13 k$\Omega$/pH. So based on equation (1) we can get the pH resolution in acid and in alkali are 0.97 pH and 0.33 pH, respectively. We believe that it reflects the change in the sensing mechanism when pH changes from acid to alkali. In an acid, the absorbed ions are primarily hydroxonium ions ($H_3O^+$), and in an alkali, the absorbed ions are primarily hydroxyl ions ($OH^-$). Indeed, it is mentioned in [26] that the hydroxyl ions may have a more ordered arrangement in the inner Helmholtz plane of the graphene/electrolyte interface than the hydroxonium ions. Furthermore, the high repeatability of multiple measurements for the particular device demonstrates that the graphene-based sensors are reusable. The reusability is probably due to the ultra-thin thickness of graphene: almost no residual can be left on the surface after the tested buffer drop is removed.

## 3.2 Sensing repeatability

The sensing repeatability of the graphene-based sensors is estimated through the measurement of three different devices with the same fabrication process as described above. All sensors demonstrate similar behaviors with resistance-time plots as shown in Figure 5a. However we have to point out that even through the sensors are fabricated under the same process, their resistance varies from device to device. Herein, the normalized resistance of different sensors is used to characterize the performance of the graphene-based sensors; it can

be defined as:

$$\frac{\Delta R}{R_r} = \frac{R - R_{min}}{R_{max} - R_{min}} \qquad (2)$$

Where $\Delta R/R_r$ is the normalized sensor resistance, $\Delta R$ is sensor resistance relative to the lowest sensor resistance, $R_r$ is the range of sensor resistance in the pH sensing test, R is the absolute value of sensor resistance, $R_{max}$ is the highest sensor resistance, and $R_{min}$ is the lowest sensor resistance.

Figure 6 shows the average values (dots) and standard deviations (error bars) of the normalized resistances at different pH values for the three sensors. All the sensors follow the same trend of pH sensitivity and the values vary in a small range. This indicates that the pH sensitivity is repeatable for multiple sensors. As the pH value increase from 4 to 10, the normalized sensor decreases proportionally. The linear relationship can be described as:

$$\frac{\Delta R}{R_r} = -0.157 \times pH + 1.68 \qquad (3)$$

3.3 Hysteresis

To test the hysteresis of the graphene-based sensors, they were exposed to pH buffers from pH-9 to pH-4 and then to pH buffers from pH-4 to pH-9. Hysteresis was observed, but the effect is in a small range. The results of a typical device are shown in Figure 7. The black line represents the testing sequence from pH-4 to pH-9 and the red line shows the reversed sequence from pH-9 back to pH-4. The two lines have a small separation but are still close to each other, and their slopes are -3.44 kΩ/pH value and -3.52 kΩ/pH, respectively.

3.4 Discussion

Interestingly, the result reported here shows an opposite trend as the result obtained from our previous experiments with carbon nanotubes [25]. However, our results are compatible with previous results on graphene research reported in the literature [26-28], where the pH sensitivities of graphene were tested with liquid-gated transistor configurations. In these studies, it has been found that there is a general trend for the conductance curve (conductance versus gate voltage) to shift towards p-doping direction with increasing pH values (along the x-axis, as illustrated in Figure 8). The fact that the annealing of graphene in air can make the graphene heavily p-doped [29, 30] implies that our conductance curve is shifted to the right; therefore, our gate-free measurements (gate voltage = 0) is on the left side of the conductance curves, making a monotonically increasing trend in the conductance measurements versus pH values (illustrated as the vertical dash line in Figure 8).

The sensing mechanism can be explained by the adsorbed ions. It is found that the adsorption processes of hydroxonium ions ($H_3O^+$) and hydroxyl ions ($OH^-$) are nonfaradaic (capacitive), meaning that the charges cannot transmit across the graphene/solution interface [26]. Therefore, according to the configuration of the electrical double layer at the graphene/electrolyte interface (as shown in Figure 9), the hydroxonium ions ($H_3O^+$) make the graphene n-doped and the hydroxyl ions ($OH^-$) make the graphene p-doped, which is consistent with other graphene studies [26, 27].

However, it has to be noted that, the sensing mechanisms of nanoscale materials including graphene could be very complicated due to many other effects, such as the effects from the substrate, Schottky barrier, gate capacitance and carrier mobility [34, 35]. Another note is that our research investigates the inherent sensing ability of graphene because our device is configured as a two-terminal resistor with a gate-free configuration. The device can also be configured as a field-effect transistor (FET) sensor with an additional terminal added to the device as a gate electrode. Recent reports investigated the pH sensitivity of graphene using a liquid-gated transistor configuration, demonstrating enhanced pH sensitivity with non-zero gate voltages [26-28]. Although their devices were manufactured based on chemical vapor deposition (CVD) and epitaxy-grown graphene samples, the measurement technology can be readily extended to our fabrication techniques with exfoliated graphene. Additional experiments will be conducted to investigate the performance of the graphene transistor-based pH sensors.

## 4. Conclusions

In summary, we have demonstrated that mechanically exfoliated graphene samples can be easily integrated into a chemical sensor with a simple mask-free process. Our sensor responds to pH changes in the surrounding liquid environment, demonstrating a linear resistance-pH relationship. We believe that the simple fabrication and miniaturized size will facilitate the applications of graphene-based sensors in a wide range of areas, especially in microfluidic sensors and point-of-care systems.

**Acknowledgements**

We thank the financialsupport from Washington State University Vancouver (WSUV) through the Undergraduate Research Mini-Grant Program(UR-MGP). Nan Lei thanks the WSUV travel grant, which made the discussion with Mingyuan Huang at Caltech possible.

Jie Xu thanks researchers from Columbia University: Changgu Lee, Changyao Chen, James Hone and Daniel Attinger for enlightening discussions.

**Figure captions:**

**Figure 1**: The mask-free fabrication process of a graphene sensor.

**Figure 2**: (a) Working principle of the focused ion beam system; (b) microscopic view of a graphene sheet connected by two platinum electrodes.

**Figure 3**: I-V curve of the device before and after annealing.

**Figure4**: Configuration of the experimental system.

**Figure5**: (a) Real-time resistance measurements of the graphene sensor when exposed to pH buffers from pH-4 to pH-10; (b) complied resistance data from multiple measurements plotted as a function of pH values.

**Figure 6:** Normalized resistances versus pH values for threegraphene-based sensors.

**Figure 7:** Hysteresis of the graphene-based sensor.

**Figure 8:** Explanation of the monotonic increasing trend in our conductivity measurements.

**Figure 9**: (a) $H_3O^+$ attached to the inner Helmholtz plane, which makes graphene "electron" doped in an acid electrolyte; (b) $OH^-$ attached to the inner Helmholtz plane, which makes graphene "hole" doped in an alkali electrolyte.

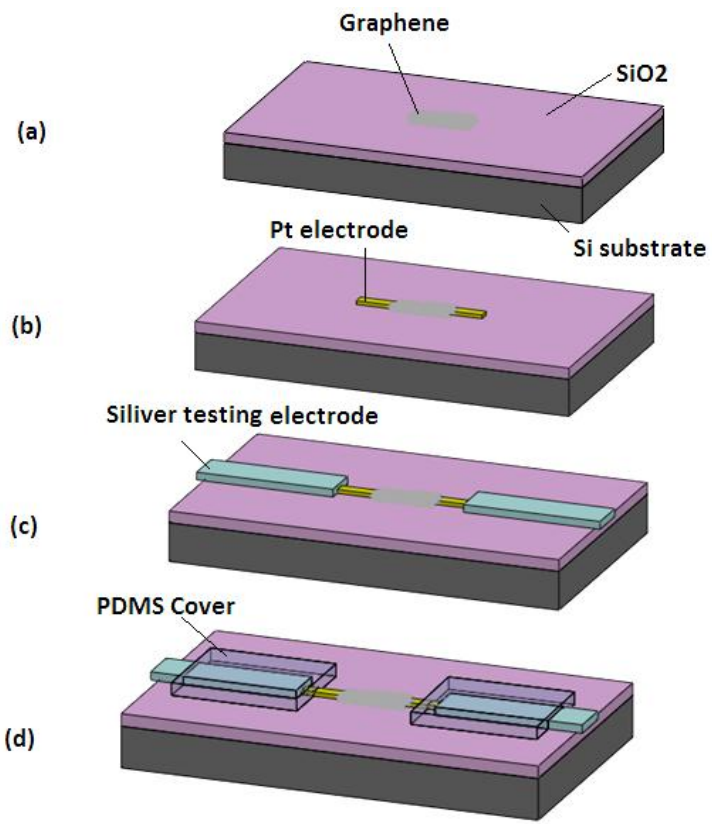

**Figure 1**: The mask-free fabrication process of a graphene sensor.

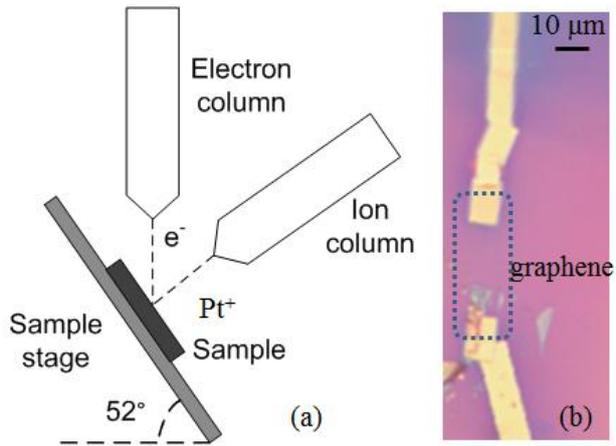

**Figure 2**: (a) Working principle of the focused ion beam system; (b) microscopic view of a graphene sheet connected by two platinum electrodes.

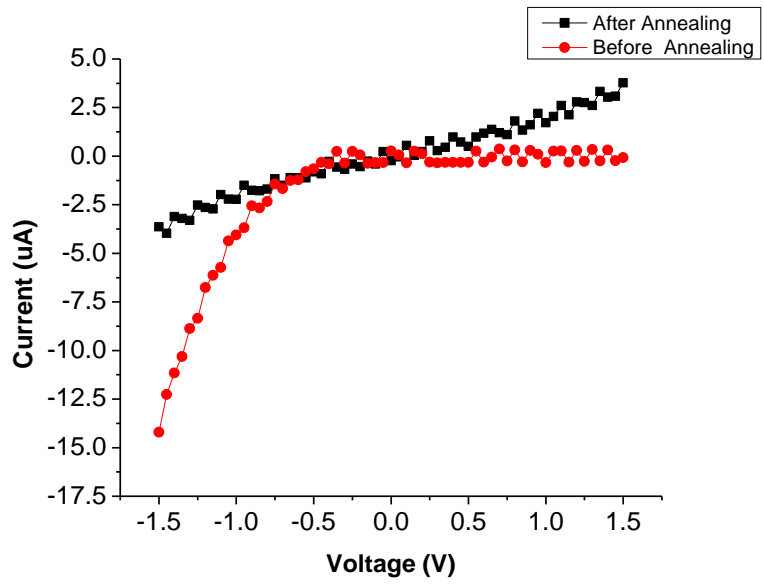

**Figure 3**: I-V curve of the device before and after annealing.

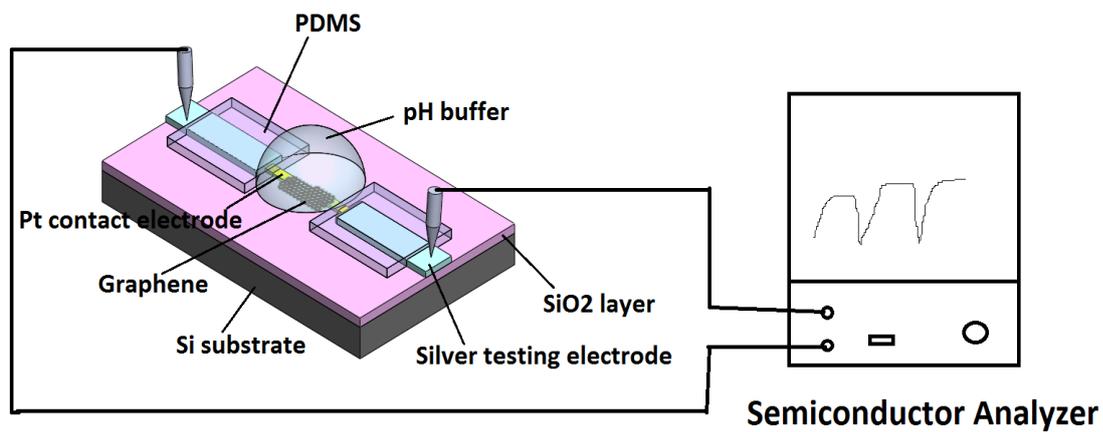

**Figure 4**: Configuration of the experimental system.

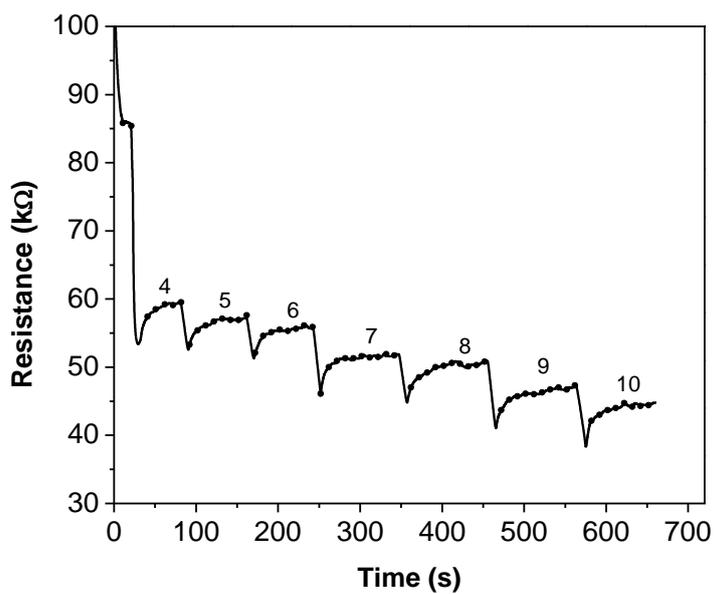

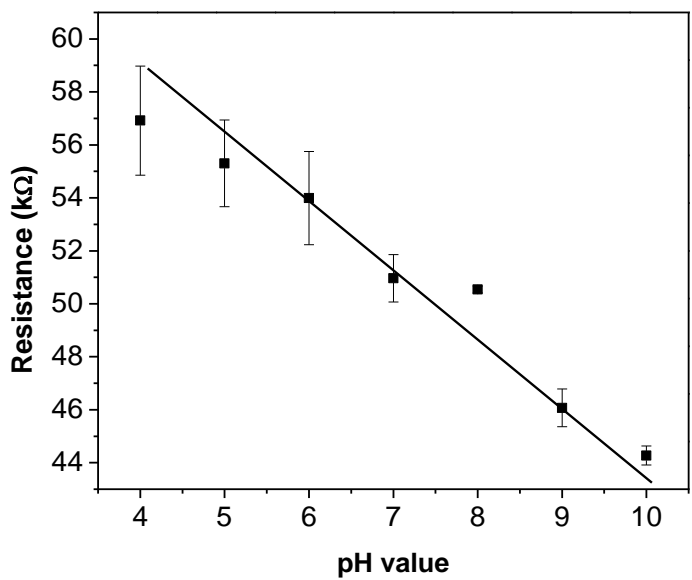

**Figure 5**: (a) Real-time resistance measurements of the graphene sensor when exposed to pH buffers from pH-4 to pH-10; (b) complied resistance data from multiple measurements plotted as a function of pH values.

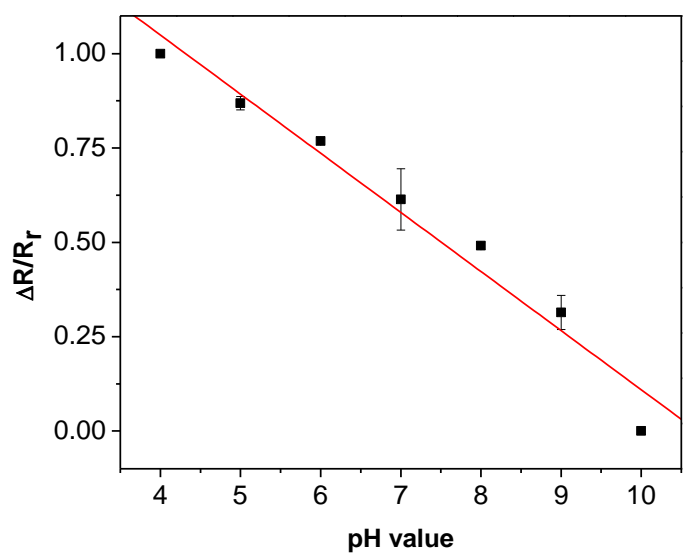

**Figure 6:** Normalized resistances versus pH values for three graphene-based sensors.

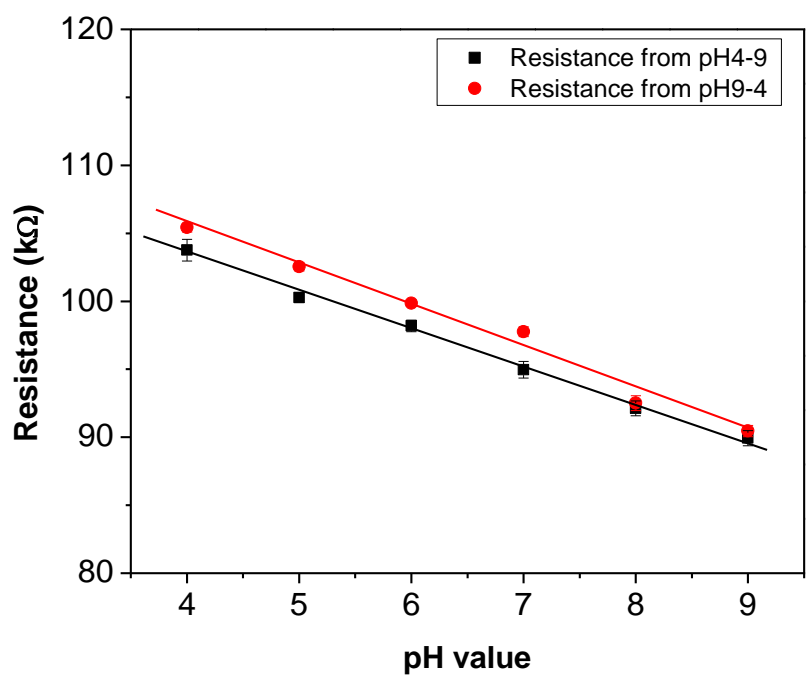

**Figure 7:** Hysteresis of the graphene-based sensor.

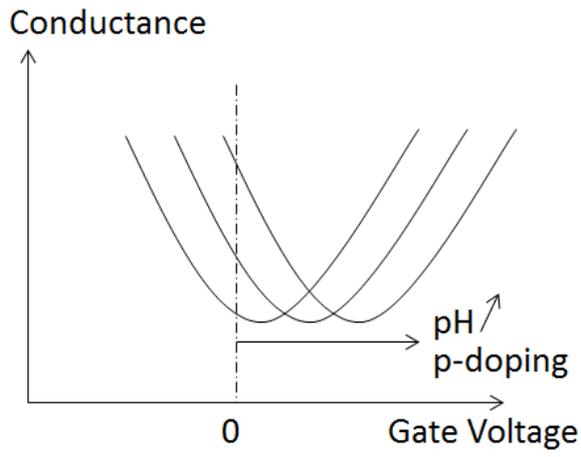

**Figure 8:** Explanation of the monotonic increasing trend in our conductivity measurements.

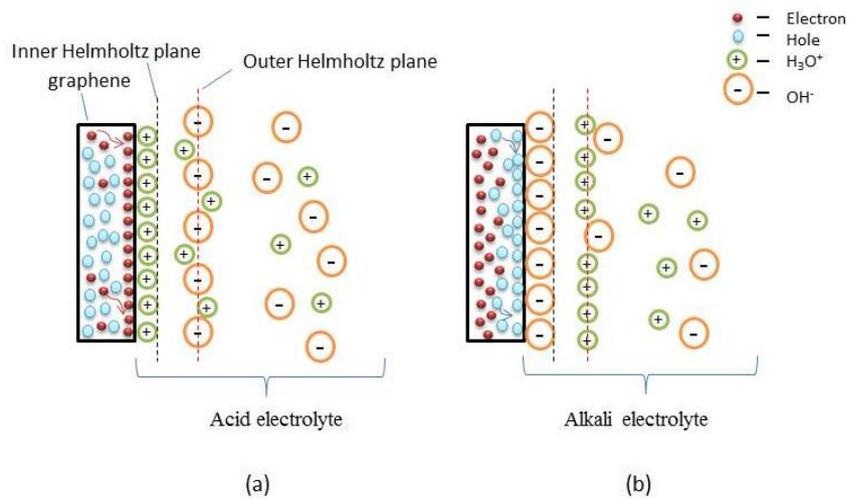

**Figure 9**: (a) $H_3O^+$ attached to the inner Helmholtz plane, which makes graphene "electron" doped in an acid electrolyte; (b) $OH^-$ attached to the inner Helmholtz plane, which makes graphene "hole" doped in an alkali electrolyte.